\documentclass{moriond}

\def\be{\begin{equation}}
\def\ee{\end{equation}}
\def\bea{\begin{eqnarray}}
\def\eea{\end{eqnarray}}

\usepackage{amsmath} 
\usepackage{amssymb}  
\usepackage{comment}
\begin{document}
\vspace*{4cm}
\title{Charm decays and $\tau$ physics at Belle and Belle II}

\author{Michele Mantovano \\
on behalf of the Belle and Belle II collaborations}

\address{Deutsche Elektronen-Synchrotron (DESY), \\\ Notkestraße 85, 22607 Hamburg, Germany}

\maketitle\abstracts{
Beyond $B$ physics, charm and $\tau$ physics constitute a central part of the Belle and Belle~II physics programs. Here, we present recent results on charm baryon decays, including first measurements and observations of several previously unmeasured modes, together with new studies in $\tau$ physics. Particular emphasis is placed on searches for CP violation in $\tau$ decays and lepton-flavor-violating processes. In this context, a search for CP violation in $\tau \to \pi K_{S} \nu_{\tau}$ decays is reported for the first time.}

\section{Charm physics}
Charm physics plays an important role at Belle and Belle~II, providing a clean environment for precision studies of weak decays and enabling detailed investigations of CP violation in charm, complementary to other experiments, while also focusing on decays of charmed baryons. In particular, recent efforts have targeted first observations and branching fraction measurements of these baryon decays, which remain comparatively less studied than mesons and offer key insights into hadronization and decay dynamics.

\subsection{First measurements of the branching fractions for the decay modes $\Xi_{c}^{0}\to \Lambda \eta$ and $\Xi_{c}^{0}\to \Lambda \eta^{\prime}$ and search for the decay $\Xi_{c}^{0}\to \Lambda \pi^{0}$ using Belle and Belle II data}

Charmed baryons provide a valuable system for studying the behavior of light quarks bound to one or more heavy quarks. Precise measurements of their branching fractions are crucial for testing theoretical frameworks and understanding their decay dynamics.
 
We report the first measurements of the branching fractions for the singly Cabibbo-suppressed decays $\Xi_c^0 \to \Lambda[\to p\pi^{-}] h^0$ ($h = \eta, \eta', \pi^0$) relative to $\Xi_c^0 \to \Xi^- \pi^+$~\cite{charm_baryon}. Theoretical expectations, derived from topological diagrams using SU(3) flavor symmetry, irreducible SU(3), or pole models, span several orders of magnitude ($\sim 10^{-5}$--$10^{-3}$), highlighting the need for experimental input to constrain these models. For instance, predictions for $\Xi_c^0 \to \Lambda \eta'$ vary from $(0.2 \pm 0.1)\times10^{-4}$ to $(16.4 \pm 10.6)\times10^{-4}$, depending on the model.

The measurement is performed using the combined Belle and Belle~II datasets, corresponding to $988.4\,\mathrm{fb}^{-1}$ and $427.9\,\mathrm{fb}^{-1}$, respectively. Unbinned extended maximum-likelihood fits are used to extract the signal. The signal shapes for the $\Lambda \eta$ and $\Lambda \eta'$ modes are described by double Gaussians, while the $\Lambda \pi^0$ mode uses bifurcated Gaussians, with combinatorial backgrounds parameterized by second-order Chebyshev polynomials. The normalization mode $\Xi_c^0 \to \Xi^- \pi^+$ is modeled with double-Gaussian signal shapes and second-order Chebyshev backgrounds. 

\begin{figure}[tb]
\centering \includegraphics[width=0.70\textwidth]{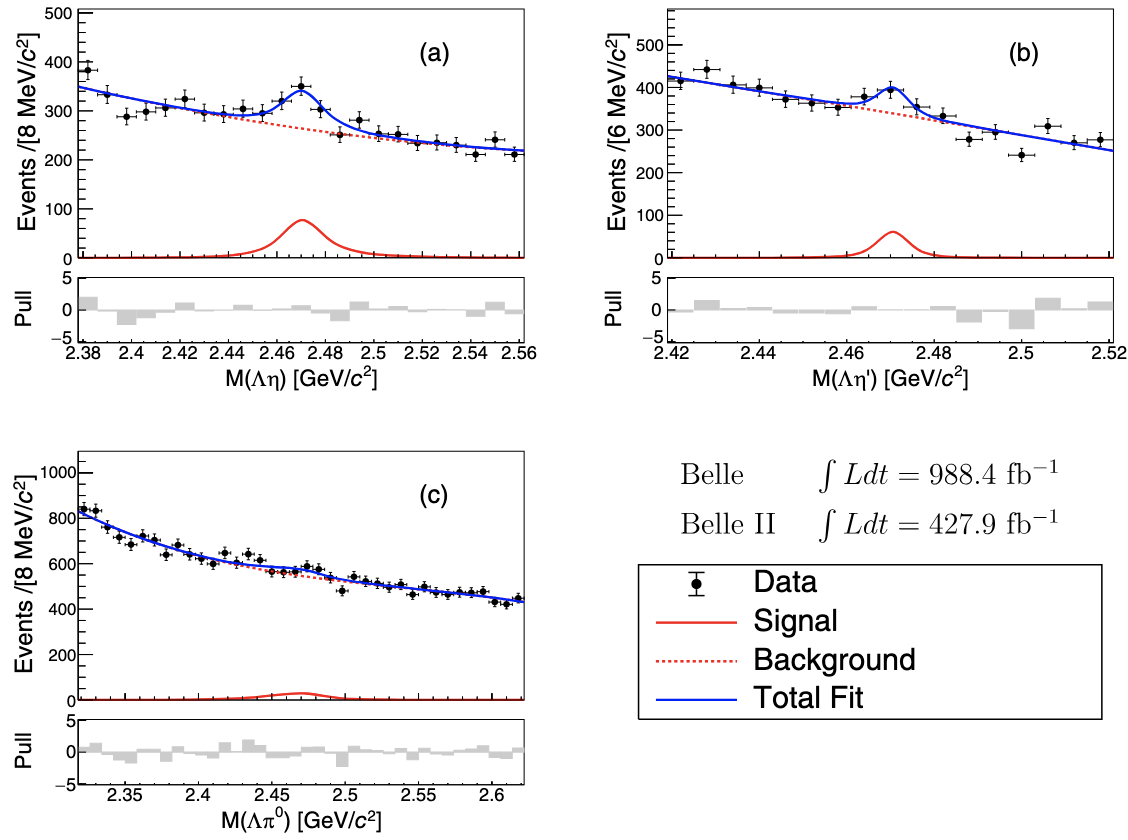}
\caption{Invariant mass distributions for (a) $\Lambda \eta$, (b) $\Lambda \eta'$, and (c) $\Lambda \pi^0$ candidates, shown together with fits to Belle and Belle~II data. Data points with error bars correspond to the observed event counts; solid red lines represent the signal probability density functions, while dashed red lines indicate the fitted combinatorial background. Overall fit results are shown by the solid blue curves, and gray histograms display the pull distributions.}
\label{fig:fit_BR_baryons_decays}
\end{figure}

Simultaneous fits across all $h^0$ decay modes and datasets, weighted by luminosity, efficiency, and secondary branching fractions, yield signal yields of $262 \pm 57$, $101 \pm 33$, and $190 \pm 120$ events for $\Lambda \eta$, $\Lambda \eta'$, and $\Lambda \pi^0$, corresponding to statistical significances of $5.3\sigma$, $3.3\sigma$, and $1.4\sigma$, respectively. After including systematic uncertainties, the significances become $5.1\sigma$ and $3.2\sigma$ for $\Lambda \eta$ and $\Lambda \eta'$, while no significant signal is observed for $\Lambda \pi^0$, resulting in a 90\% CL upper limit on the signal yield of 454 events. Fit projections for all three modes are shown in Fig.~\ref{fig:fit_BR_baryons_decays}.

The corresponding absolute branching fractions are
\begin{align}
\mathcal{B}(\Xi_c^0 \to \Lambda \eta) &= (5.95 \pm 1.30 \pm 0.32 \pm 1.13)\times 10^{-4}, \nonumber \\
\mathcal{B}(\Xi_c^0 \to \Lambda \eta') &= (3.55 \pm 1.17 \pm 0.17 \pm 0.68)\times 10^{-4},  \\
\mathcal{B}(\Xi_c^0 \to \Lambda \pi^0) &< 5.2 \times 10^{-4} \quad \text{at 90\% CL}, \nonumber 
\end{align}
where the uncertainties are statistical, systematic, and due to $\mathcal{B}(\Xi_c^0 \to \Xi^- \pi^+)$, respectively. These results are consistent with most theoretical predictions, with SU(3)$_F$-based calculations~\cite{SU3_prediction} agreeing within $1\sigma$ for $\Lambda \eta$ and $\Lambda \eta'$, and all predictions lying below the $\Lambda \pi^0$ upper limit. The branching fraction ratios, being independent of the absolute scale of $\mathcal{B}(\Xi_c^0 \to \Xi^- \pi^+)$, provide a model-independent test of theoretical models of charmed baryon decays.\\

\subsection{Observation of the radiative decay $D_{s}(2317)^{+}\to D_{s}^{*}\gamma$ using Belle and Belle II data}

\begin{figure}[tb]
\centering 
\includegraphics[width=0.45\textwidth]{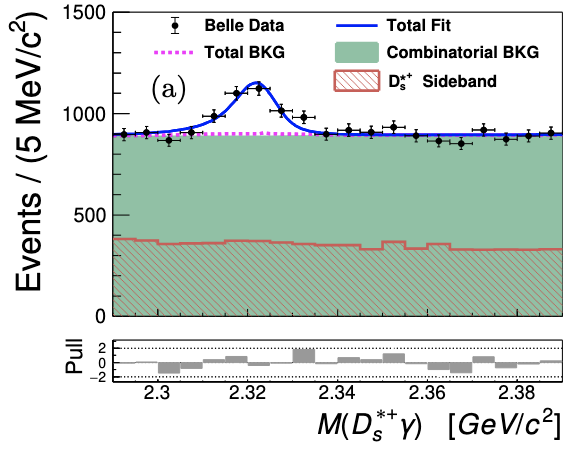}
\includegraphics[width=0.458\textwidth]{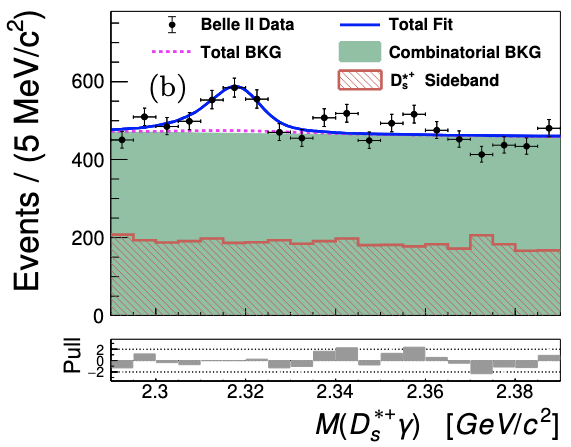} 
\caption{Simultaneous fits to the $M(D^{*+}_s \gamma)$ spectra from (a) Belle and (b) Belle II datasets. The blue curves show the overall fit to the signal, while the violet lines correspond to the fitted total background. The green shaded regions indicate the combinatorial background component, and the red-slash histograms display the normalized $D^{*+}_s$ sideband contribution.}
\label{fig:fit_Dsgamma}
\end{figure}

The study of exotic hadrons provides key insights into non-perturbative QCD. The charm-strange mesons $D_{s0}^*(2317)^+$ and $D_{s1}(2460)^+$ are of particular interest, as their masses lie below the quark-model predictions for $c\bar{s}$ mesons. Various interpretations have been proposed, including molecular, tetraquark, conventional quark–antiquark, and mixed configurations, but their nature remains unresolved. The $D_{s0}^*(2317)^+$ was first observed in $D_s^+\pi^0$ decays by BaBar~\cite{BaBar_radiative} and later confirmed by CLEO~\cite{CLEO_radiative} and Belle~\cite{Belle_radiative}. 
Radiative decays, such as $D_{s0}^*(2317)^+ \to D_s^{*+}\gamma$, are sensitive probes of its internal structure; however, previous searches by CLEO~\cite{CLEO_radiative}, Belle~\cite{Belle_radiative}, and BaBar~\cite{BaBar_radiative_2} found no signal, with the most stringent upper limit on the branching fraction ratio, 5.9\% at 90\% CL, set by CLEO~\cite{CLEO_radiative}.

\indent We report the first observation of $D_{s0}^*(2317)^+ \to D_s^{*+}\gamma$ with $D_s^{*+} \to D_s^+\gamma$, measuring its rate relative to $D_{s0}^*(2317)^+ \to D_s^+\pi^0$~\cite{observation_radiative}. The analysis uses $980.4~\mathrm{fb}^{-1}$ of Belle data collected near $\Upsilon(nS)$ ($n=1\!-\!5$) and $427.9~\mathrm{fb}^{-1}$ of Belle~II data collected near $\Upsilon(4S)$ and at 10.75~GeV. The $D_s^+$ mesons are reconstructed in the $\phi\pi^+$ and $K^+\bar{K}^{*0}$ modes. Using the hadronic decay as a reference cancels several systematic uncertainties in the branching fraction ratio, providing crucial input to discriminate among theoretical models.

\indent The $D_{s0}^*(2317)^+ \to D_s^{*+}\gamma$ signal is clearly observed in the $M(D_s^{*+}\gamma)$ spectra for both Belle and Belle~II datasets. Potential peaking backgrounds from $D_{s0}^*(2317)^+ \to D_s^+\pi^0$, $D_{s1}(2460)^+ \to D_s^{*+}\pi^0$, and misassociated photons are small and are modeled using simulation.

\indent The branching fraction ratio $R = \mathcal{B}(D_{s0}^*(2317)^+ \to D_s^{*+}\gamma)/\mathcal{B}(D_{s0}^*(2317)^+ \to D_s^+\pi^0)$
is extracted via a simultaneous unbinned extended maximum-likelihood fit to the Belle and Belle~II spectra. Signal shapes are described by Crystal Ball functions convolved with triple-Gaussian functions, while combinatorial backgrounds are modeled with $1^{st}$-order polynomials. 
Fit projections are shown in Fig.~\ref{fig:fit_Dsgamma}.

\indent Based on the combined datasets, this represents the first observation of the radiative decay $D_{s0}^*(2317)^+ \to D_s^{*+}\gamma$ in $e^+e^- \to c\bar c$, with a significance exceeding $10\sigma$. The measured branching fraction ratio is 
\begin{equation}
R = [7.14 \pm 0.70~\text{(stat.)} \pm 0.23~\text{(syst.)}]\%,
\end{equation}
which is larger than predictions favoring a molecular interpretation but consistent with some models assuming a conventional $c\bar s$ structure. This suggests that $D_{s0}^*(2317)^+$ may be a mixture of a pure $c\bar s$ state and a molecular component.

\section{$\tau$ physics}
At Belle and Belle~II, $\tau$-pair events are produced with a large cross section in a clean experimental environment, making these experiments not only $B$-factories but also genuine $\tau$-factories. However, the presence of one or more neutrinos in the final state prevents a full kinematic reconstruction of $\tau$ decays. To identify these events, the thrust axis is used to characterize the event topology. In the center-of-mass frame, the event is divided into two hemispheres by a plane perpendicular to the thrust axis, naturally separating the two $\tau$ decays. Events are then classified according to the charged-particle multiplicity in each hemisphere, enabling efficient background suppression and robust reconstruction of $\tau$ decay modes.

\subsection{Search for CP violation in the $\tau \to \pi K_{S}^{0}\nu_{\tau}$ decays using Belle II data}

In the Standard Model (SM), CP violation arises from a complex phase in the Cabibbo--Kobayashi--Maskawa matrix and has been observed in mesons, as well as in beauty baryons. However, its magnitude is insufficient to explain the matter--antimatter asymmetry, motivating searches in other sectors, such as charged leptons. In $\tau$ decays involving a $K_S^0$ in the final state, a CP-violating decay-rate asymmetry can arise from $K^0$--$\bar{K}^0$ mixing, with an SM prediction of $A_{CP} \simeq 0.33\%$~\cite{SM_CPV_tau}. Any significant deviation would indicate physics beyond the SM. Previous measurements by BaBar~\cite{BaBar_CP_tau} showed a 2.82$\sigma$ tension with the SM.

We present here for the first time a measurement of $A_{CP}$ in $\tau^- \to \pi^- K_S^0 (\geq 0\pi^0)\nu_\tau$ using Belle~II data corresponding to $362~\mathrm{fb}^{-1}$, where one $\tau$ decays hadronically and the other leptonically. After background subtraction, the raw asymmetry is defined as
\begin{equation}
A_{\mathrm{raw}} =
\frac{N(\tau^+ \to h^+ K_S^0 (\geq 0\pi^0)\bar{\nu}_\tau) - 
      N(\tau^- \to h^- K_S^0 (\geq 0\pi^0)\nu_\tau)}
     {N(\tau^+ \to h^+ K_S^0 (\geq 0\pi^0)\bar{\nu}_\tau) + 
      N(\tau^- \to h^- K_S^0 (\geq 0\pi^0)\nu_\tau)}.
\end{equation}

This observable includes contributions from several decay modes and is not a purely CP-violating quantity, as it is affected by production and detector-induced asymmetries. These include forward--backward production asymmetry, reconstruction asymmetries of charged hadrons and leptons, and trigger-induced effects.
These contributions are corrected using a control sample of $\tau^- \to h^- h^+ h^- \nu_\tau$ decays, which shares similar kinematic properties but does not exhibit intrinsic CP violation. In addition, the decay $\tau^\pm \to \pi^\pm K_S^0 \nu_\tau$ is also affected by $K^0$–$\bar{K}^0$ mixing, CP violation, and differing interactions with detector material. This effect is modeled using the time evolution of the neutral kaon system in matter, with the dominant correction arising from kaon absorption, while interference contributions are found to be negligible. 

The CP-violating asymmetry is then obtained as

\begin{figure}[tb]
\centering 
\includegraphics[width=0.45\textwidth]{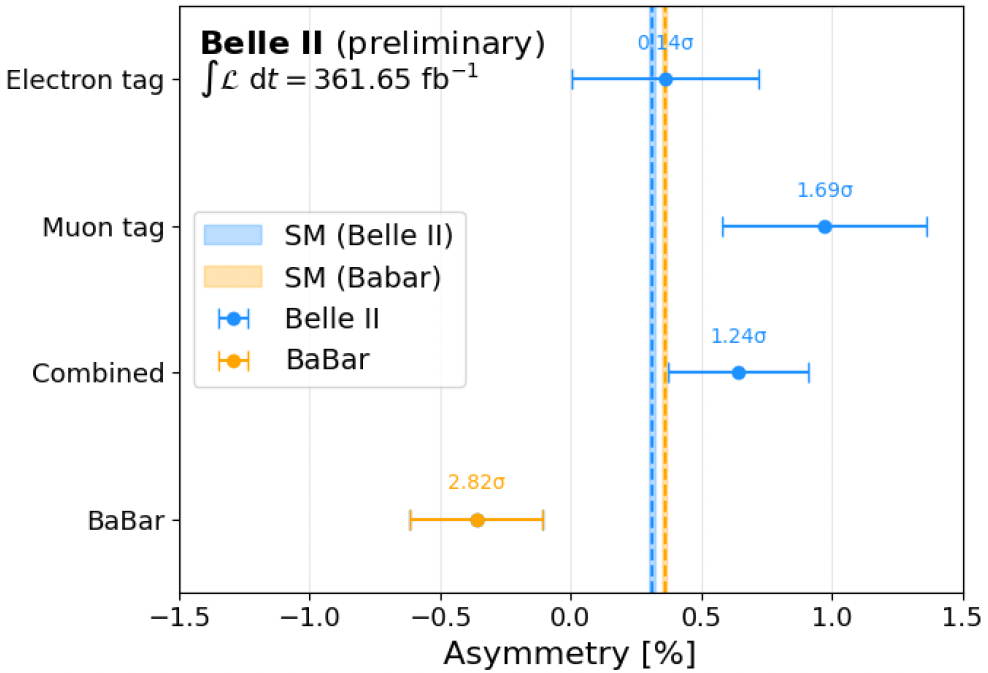}
\includegraphics[width=0.46\textwidth]{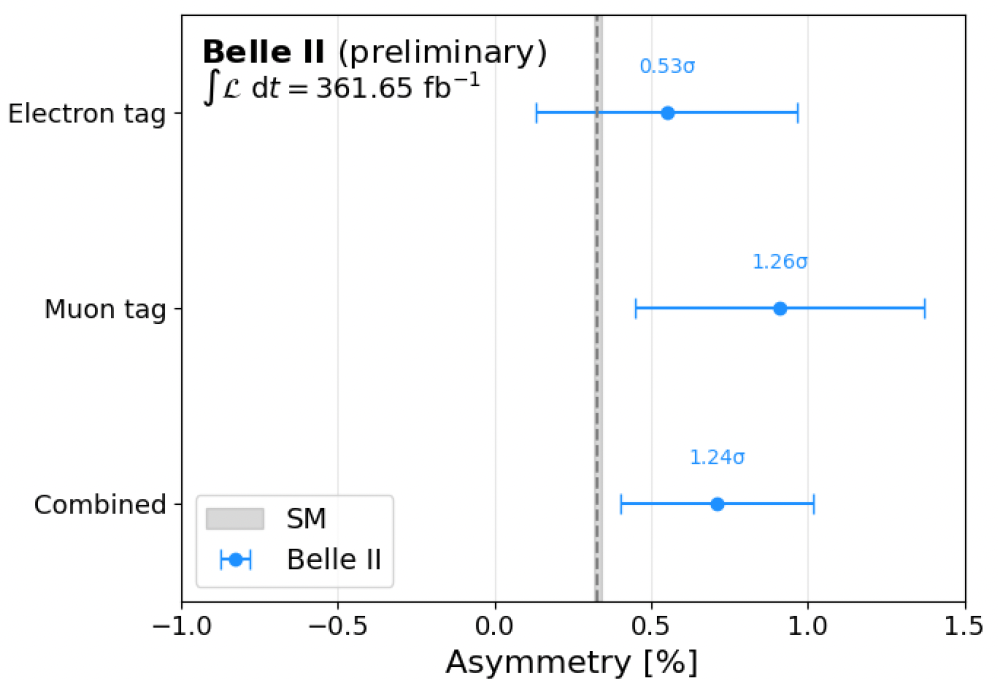} 
%\caption{(Left) Measured CP asymmetry in $\tau^- \to \pi^- K_S^0 \nu_\tau$ decays for electron-tagged, muon-tagged, and combined Belle~II samples, compared with the previous BaBar measurement~\cite{BaBar_CP_tau}. The SM expectations, corrected for $K_S^0$ selection efficiency, are shown as dashed lines. (Right) Efficiency-corrected CP asymmetry for the same decays, showing electron-tagged, muon-tagged, and combined Belle~II results. The SM prediction, $A_{CP}^{\mathrm{SM}} = 0.33\%$, is indicated by a dashed gray line.}
\caption{(Left) Measured CP asymmetry in $\tau^- \to \pi^- K_S^0 \nu_\tau$ decays for electron-tagged, muon-tagged, and combined Belle~II samples, compared with the previous BaBar measurement. The SM expectations, corrected for $K_S^0$ selection efficiency, are shown as dashed lines. (Right) Efficiency-corrected CP asymmetry for the same decays, showing electron-tagged, muon-tagged, and combined Belle~II results. The SM prediction, $A_{CP}^{\mathrm{SM}} = 0.33\%$, is indicated by a dashed gray line.}
\label{fig:ACP}
\end{figure}
\begin{equation}
A_{CP} = A_{\mathrm{raw}} - A_{\mathrm{corr}},
\end{equation}
where $A_{\mathrm{corr}}$ accounts for all non-CP contributions. The measured asymmetry receives contributions from different decay channels and can be expressed as a weighted average of their individual asymmetries.  

The asymmetry is further affected by the decay-time dependence of the $K_S^0$ reconstruction efficiency, which modifies the relative contributions of $K^0$ and $\bar{K}^0$. A consistent comparison between the measurement and the SM prediction therefore requires accounting for this effect.

The measured asymmetry after applying the decay-time–efficiency correction is
\begin{equation}
A_{CP} = (0.71 \pm 0.26 \pm 0.06 \pm 0.15)\%,
\end{equation}
where the uncertainties are statistical, systematic, and due to the efficiency correction, respectively. This result, consistent with the SM prediction of $A_{CP}^{\mathrm{SM}} \simeq 0.33\%$ within $1.24\sigma$, shows no evidence for additional sources of CP violation and does not confirm the previously reported tension by BaBar, as shown in Fig.~\ref{fig:ACP}. With a precision already comparable to the BaBar measurement, it demonstrates the strong potential of Belle~II to provide a decisive test of CP violation in $\tau$ decays as larger data samples become available.

\subsection{Search for lepton-flavor violation  $\tau \to \ell \eta$ decays using Belle II data}

Lepton flavor violation (LFV) is forbidden in the SM with massless neutrinos, but neutrino oscillations imply that it can occur via neutrino exchange at an unobservable level ($\mathcal{O}(10^{-50})$). Many beyond-SM scenarios predict enhanced branching fractions for LFV $\tau$ decays, making them accessible at current high-luminosity experiments. Representative processes include $\tau^- \to \ell^- \eta$ ($\ell = e, \mu$), with models such as leptoquark, the Littlest Higgs model with $T$-parity, and type I--III seesaw predicting branching fractions $\mathcal{O}(10^{-8})$, within reach of Belle~II. Previous searches by CLEO~\cite{CLEO_leta}, BaBar~\cite{BaBar_leta}, and Belle~\cite{Belle_leta} found no signal, 
with Belle setting the most stringent limits of $9.2\times 10^{-8}$ ($6.5\times 10^{-8}$) for the electron (muon) channel at 90\% CL.

We present a search for $\tau \to \ell \eta$ decays using $427.9~\mathrm{fb}^{-1}$ of Belle~II data. Signal candidates are reconstructed via $\tau^- \to \ell^- \eta$ ($\eta\to \gamma\gamma, \pi^+\pi^-\pi^0$), with the tag-side $\tau^+$ decaying in a one-prong mode. The signal and background regions are defined in the $(M_{\tau}, \Delta E)$ plane, where the signal peaks at $M_{\tau}\simeq 1.777~\mathrm{GeV}/c^2$ and $\Delta E\simeq 0$. To define the analysis regions, the parameters of the signal distributions in $M_{\tau}$ and $\Delta E$ are determined from fits using two asymmetric Crystal Ball functions. Several regions are defined, including a signal region optimized via a figure-of-merit, an extended-signal region (containing 90\% of the signal), sidebands, and an outer $\pm 20\delta$ box.

\begin{figure}[tb]
\centering 
\includegraphics[width=0.80\textwidth]{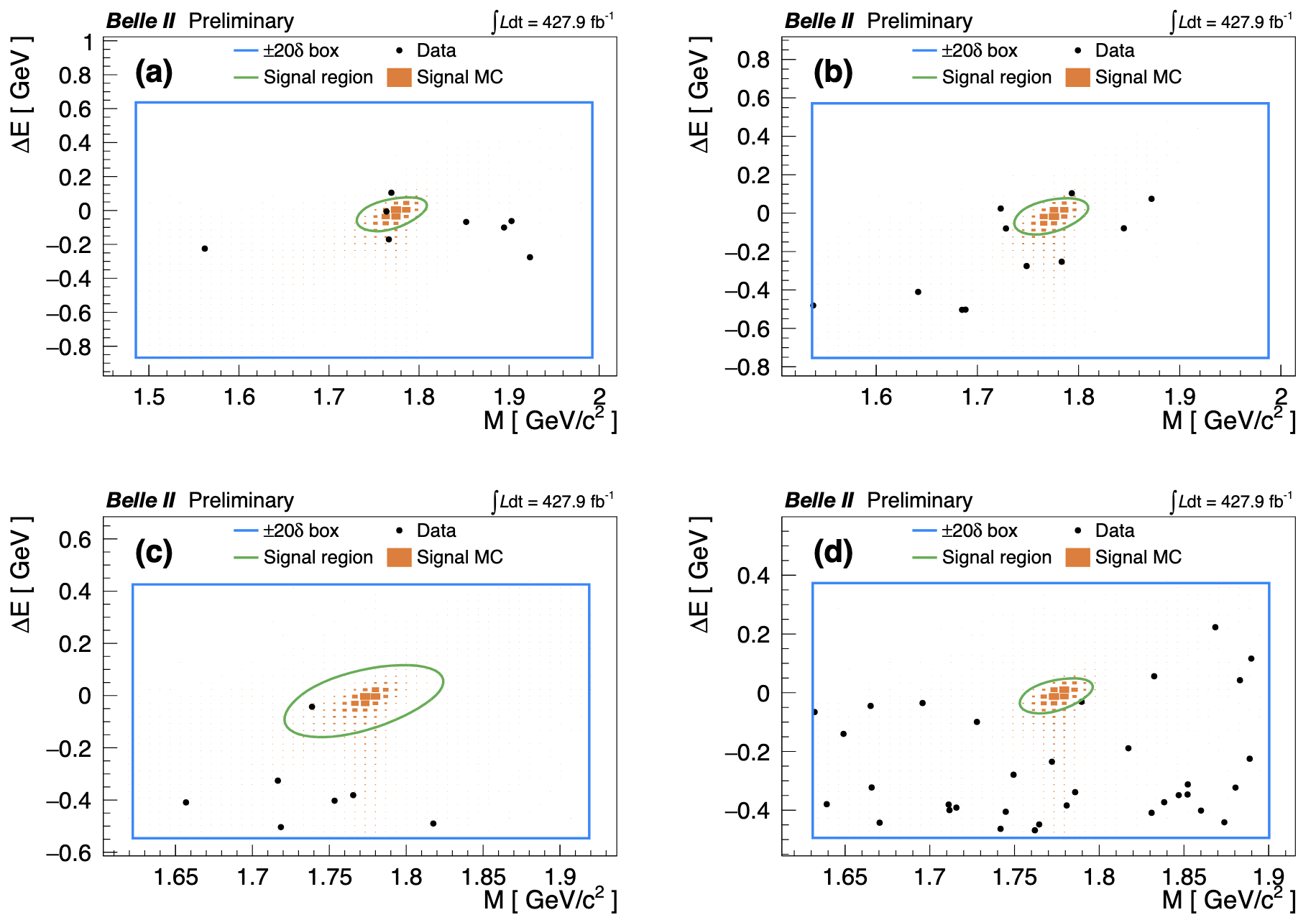}
\caption{Event distributions for $\tau^- \to \ell^- \eta$ in the $(M_{\tau}, \Delta E)$ plane: (a) $e^-\eta(\to \gamma\gamma)$, (b) $\mu^-\eta(\to \gamma\gamma)$, (c) $e^-\eta(\to \pi^+\pi^-\pi^0)$, and (d) $\mu^-\eta(\to \pi^+\pi^-\pi^0)$. The analysis region is indicated by the blue box, while the signal region is highlighted with a green ellipse. Simulated signal events are shown as filled orange boxes proportional to the event density; observed data are marked as black points.}
\label{fig:tau2leta}
\end{figure}

Background suppression depends on the decay modes. In the electron channel, Bhabha events are reduced using kinematic cuts on track opening angles and electromagnetic calorimeter energies, whereas no equivalent suppression is applied in the muon channel. For $\eta\to\gamma\gamma$, both cut-based and multivariate selections are used due to higher background, whereas $\eta\to\pi^+\pi^-\pi^0$ relies on cut-based methods alone. The expected background in the signal region is estimated using data-driven techniques: a counting method with nearly uncorrelated variables for $\eta\to\gamma\gamma$, and a two-dimensional unbinned maximum-likelihood fit in $(M_{\tau}, \Delta E)$ for $\eta\to\pi^+\pi^-\pi^0$.

\indent After full selection, one event is observed in the $e^-\eta$ channels ($\eta\to\gamma\gamma, \pi^+\pi^-\pi^0$) and none in the corresponding $\mu^-\eta$ channels. 
The number of signal events for each channel, together with the definition of the signal region, is shown in Fig.~\ref{fig:tau2leta}.
Upper limits at 90\% CL are determined using the frequentist CL$_s$ method with profile likelihoods and pseudo-experiments, incorporating systematic uncertainties. The observed limits are
\begin{equation}
\mathcal{B}(\tau^-\to e^-\eta) < 9.21\times10^{-8} \quad \text{and} \quad
\mathcal{B}(\tau^-\to\mu^-\eta) < 4.23\times10^{-8}.
\end{equation}

\indent The results show no evidence for LFV, agree with the previous best limit from Belle for the electron mode, and set the most stringent constraint so far on the muon mode. With increasing Belle~II luminosity, statistical precision will improve, enhancing sensitivity to LFV $\tau$ decays.

\section{Conclusion}
The Belle and Belle~II experiments provide an exceptional environment for precision studies of beauty, charm, and $\tau$ physics. Their clean $e^+e^-$ environment and large integrated luminosities enable precise measurements of branching fractions in charmed baryon decays, improving our understanding of weak decay dynamics and light-quark interactions. They also offer high-precision measurements of CP violation and mixing in charm, along with competitive sensitivity in $\tau$ CP-violation studies. In addition, Belle and Belle~II lead searches for lepton-flavor-violating $\tau$ decays, setting world-leading limits that constrain new physics scenarios. Overall, large datasets and excellent detector performance ensure a leading role in flavor physics, providing both stringent SM tests and sensitive probes of rare or forbidden processes.

\section*{References}
\bibliography{moriond}

\end{document}